\documentclass[prl,aps,amsmath,nofootinbib,amssymb,showkeys,twocolumn]{revtex4-1}
\pdfoutput=1
\usepackage{verbatim,graphics,graphicx,color,slashed,textcomp}
\usepackage{ulem} 
\usepackage[colorlinks=true,
            linkcolor=red,
            urlcolor=blue,
            citecolor=blue]{hyperref}

\graphicspath{{figures/}}

\newcommand{\eV}{\mathrm{eV}}

\newcommand{\MeV}{\mathrm{MeV}}
\newcommand{\GeV}{\mathrm{GeV}}
\newcommand{\TeV}{\mathrm{TeV}}

\newcommand{\eff}{\textrm{eff}}
\newcommand{\planck}{\texttt{Planck}}

\begin{document}

\title{Light dark photon and fermionic dark radiation  for \\ the Hubble constant and 
the structure formation}
\author{P. Ko and Yong Tang}
\affiliation{School of Physics, Korea Institute for Advanced Study, Seoul 02455, South Korea}
\date{\today}

\begin{abstract}
Motivated by the tensions in the Hubble constant $H_0$ and the structure growth $\sigma_8$ 
between \planck ~results and other low redshift measurements, we discuss some cosmological 
effects of a dark sector model in which dark matter (DM) interacts with  fermionic
dark radiation (DR)  through a light gauge boson (dark photon). Such kind of models are very 
generic in particle physics with a dark sector with  dark gauge symmetries. 
The effective number of neutrinos is increased by $\delta N_\eff \sim 0.5$ 
 due to light dark photon and fermionic DR, thereby resolving
the conflicts in $H_0$. The elastic scattering between DM 
and DR induces suppression for DM's density perturbation, but without acoustic oscillations. 
For weakly-interacting DM around $100\GeV$, the new gauge coupling should be 
$\sim 10^{-4}$ to have sizable effect on matter power spectrum in order to relax the tension 
in $\sigma_8$. 

\end{abstract}
\maketitle

\section{Introduction}
It has been established that about $83\%$ of matter content ($\Omega_m=0.3065\pm 0.0072$)~\cite{Planck:2015xua} in our universe is composed of dark matter (DM). The standard cold DM (CDM) together with cosmological constant $\Lambda$, $\Lambda$CDM model, is very compelling and convincing to explain our current observations. Despite of this remarkable success, we are 
still struggling to disentangle the particle identities of DM since all the confirmed evidence for DM  
come from gravitational interaction of DM. Any unexpected signatures in astrophysics, cosmology 
and particle physics may help us to better understand particle physics nature of DM. 

Meanwhile, there are still some persistent tensions in the measurement of the Hubble constant 
$H_0$ and the structure growth rate $\sigma _8$ (the amplitude of matter fluctuations at scale 
around 8 Mpc). The latest analysis~\cite{Riess:2016jrr} of Hubble Space Telescope (HST) data 
gives $H_0=73.24\pm 1.74$km s$^{-1}$Mpc$^{-1}$, which is about 3.4$\sigma$ higher than the 
value given by \planck~\cite{Planck:2015xua} within the $\Lambda$CDM model.  
Also, \planck ~data yields $\sigma _8=0.815\pm 0.009$ which is relatively larger than the low  
redshift measurements, such as weak lensing survey CFHTLenS~\cite{Heymans:2012gg}, 
$\sigma_8(\Omega_m/0.27)^{0.46}=0.774\pm 0.040$. 

The above tensions could be due to systematic uncertainties, or they may indicate new physics model beyond the standard $\Lambda$CDM. For example, increasing the effective number of neutrinos by $\delta N_\eff \simeq 0.4-1$ with dark radiation (DR) could resolve the conflict 
between \planck ~and HST data~\cite{Riess:2016jrr}, which, however, unfortunately would give 
an even larger $\sigma_8$. Or it is possible to extend the six-parameter $\Lambda$CDM with 
varying dark energy, dark matter, neutrino mass, running spectral index, and so on~\cite{Pourtsidou:2016ico, DiValentino:2016hlg, Qing-Guo:2016ykt, Archidiacono:2016kkh, Wyman:2013lza, Zhang:2014lfa, Lesgourgues:2015wza}, to relax these tensions in $H_0$ and $\sigma_8$. 

In this letter, we shall explore a dark sector model in which DM interacts with DR through light 
dark photon and address the above issues.  The interaction between DM and DR  causes a 
suppression of the matter power spectrum through diffusion or collisional damping which can 
give a smaller $\sigma _8$.   Also the natural presence of DR would relieve the tension between 
HST and \planck.

This paper is organized as following. Firstly, we shall introduce our model setup with the 
conventions and the relevant parameters. Then we  discuss the corresponding phenomenologies, 
DM relic density, prediction of $\delta N_\eff$ and the DM-DR scattering with late kinetic decoupling. 
Later, we show some numerical results on the matter power spectrum. 
Finally, we give our summary.

\section{The Model}\label{sec:intro}
We introduce a dark sector with a new $U(1)$ dark gauge symmetry and coupling $g_X$, dark 
photon field $V_\mu$, scalar $\Phi$, massive fermion $\chi$ (DM) and massless $\psi$ (DR). 
All these new fields are living in the dark sector, thereby being SM gauge singlets. We assign 
$U(1)$ charges $q_f=1,2,2$ to $\chi, \psi,\Phi$, respectively. Then the general gauge invariant 
Lagrangian is 
\begin{align}\label{eq:lagrangian}
\mathcal{L} 
=&- \frac{1}{4}V_{\mu\nu}V^{\mu\nu}+ D_{\mu} \Phi^{\dagger}D^{\mu}\Phi+\bar{\chi}\left(i\slashed{D}-m_{\chi}\right)\chi
+\bar{\psi}i\slashed{D}\psi \nonumber\\
& - \left(y_\chi \Phi^\dagger\bar{\chi}^{c}\chi+ y_\psi \Phi \bar{\psi}N+h.c.\right)-V(\Phi , H),
\end{align} 
where $N$ is the singlet right-handed (RH) neutrino which couples to the left-handed 
(LH) neutrinos in the SM through usual Yukawa terms, the superscript `$c$' stands for charge conjugate, the covariant derivative is defined as $ D_\mu f = (\partial_\mu - iq_f g_X V_\mu)f$, $\slashed{D}\equiv \gamma^\mu D_\mu$ and $V_{\mu\nu}=\partial_\mu V_\nu - \partial_\nu V_\mu$. Note that $\Phi$ does not develop a vacuum expectation value (VEV) and this new $U(1)$ is a good symmetry. We could introduce a mass for $V_\mu$ and possible gauge kinetic mixing term \footnote{This could be achieved by a nonzero VEV of $\Phi$, or by introducing another $U(1)$-charged scalar with nonzero VEV. See Ref.~\cite{Essig:2013lka} for constraints on kinetic mixing.}, which however is not essential for our discussions, and which we shall come back to later. 

Except for the Higgs and Yukawa terms, our model is very similar to the structure in standard model. Some simple variants of this model is equally suited for our interests in the paper. 
For example, $\Phi$ can be a singlet and couples as 
$y_\chi \Phi\bar{\chi}\chi+ y_\psi \Phi \bar{\psi}\psi$.
In any case $\Phi$ is not stable and can decay into $\psi$, and $\psi$ can be thermalized with $\chi,\Phi$ and $V_\mu$
through the Yukawa couplings with $\Phi$.

We note that a similar setup was discussed in Ref.~\cite{Lesgourgues:2015wza}, 
where the authors assumed $q_\chi=1\neq q_\psi$, but did not consider possible Yukawa 
interactions  between $\psi$ with the $\Phi$.  
Yukawa interaction among $\Phi$ and $\psi$ can lead to thermalization of $\psi$ at high temperature, which is
different from thermalization mechanism at lower temperature through dark gauge interactions considered in 
Ref.~\cite{Lesgourgues:2015wza}, and the resulting $\delta N_{\rm eff}$ would be different.

Finally, the connection to the SM sector can be established in a straightforward manner 
through the Higgs portal term, 
$V\supset\lambda_{\Phi H}\Phi^\dagger\Phi H^\dagger H$, where $H$ is the SM Higgs doublet. 
Simple estimation shows that $\Phi$ and dark sector can be in thermal equilibrium with SM particles when the Universe is around $\TeV$ if $|\lambda_{\Phi H}|\gtrsim 10^{-6}$.

\section{Phenomenology}\label{sec:constraint}

Now, let us discuss some relevant phenomenology and constraints, based on the Lagrangian 
of Eq.~(\ref{eq:lagrangian}). 

{\it Relic density}: For thermal DM $\chi$ and $m_\chi > m_\Phi$, its relic abundance is mostly determined by the annihilation process $\chi+\bar{\chi}\rightarrow \Phi +\Phi^\dagger$. At tree-level approximation, we have the thermal cross section
\begin{equation}
\langle \sigma v\rangle \sim \frac{y_\chi^4}{16\pi m^2_\chi},
\end{equation}
and the total relic density of $\chi$ and $\bar{\chi}$ would be
\begin{equation}\label{eq:relic}
\Omega h^2\simeq 0.1\times\left(\frac{y_\chi}{0.7}\right)^{-4}\left(\frac{m_\chi}{\TeV}\right)^2. 
\end{equation}

The value of $y_\chi$ determined by Eq.~(\ref{eq:relic}) can be treated as the upper limit for 
$y_\chi$, since if there were other annihilation processes contributing to the depletion of $\chi$ particles, then $y_\chi$ could be smaller. For instance, $\chi+\bar{\chi}\rightarrow \psi +\bar{\psi}$ can be important if $y_\psi>y_\chi$. However, for our interests in this paper, the qualitative relation above between $y_\chi$ and $m_\chi$ would be sufficient, which means that for $\TeV$-scale $\chi$ it is expected to have $y_\chi\sim 0.7$ to get the correct relic density. 

DM $\chi$'s self-scattering through exchanging $\Phi$ can be sizable if the mass of $\Phi$ 
($m_\Phi$) is small, which is the central topics in recent self-interacting dark matter scenarios, 
(see Refs.~\cite{Aarssen:2012fx, Ko:2014nha, Bringmann:2013vra, Ko:2014bka, Chu:2014lja, Nelson:2014mva, Chu:2015ipa, Arhrib:2015dez, Kang:2015aqa, Kainulainen:2015sva, Bernal:2015bla, Choi:2015bya, Ma:2015roa, Chacko:2015noa, Baek:2013dwa, Buen-Abad:2015ova, Heikinheimo:2015kra, Arhrib:2015dez, Chang:2016pya, Ko:2014lsa, Ko:2015nma, Cline:2013zca, Foot:2004wz, CyrRacine:2012fz, Boddy:2014yra, Kang:2016xrm, Tang:2016mot,Feng:2009hw, Buckley:2009in, Loeb:2010gj, Tulin:2013teo, Petraki:2014uza, Buckley:2014hja, DelNobile:2015uua, ethos, Bernal:2015ova, Binder:2016pnr, Bringmann:2016ilk, Geng:2013oda} for examples). 
In general, for $\mathcal{O}(100 \GeV)$ DM $\chi$,  $\Phi$ with $m_\Phi\sim \mathcal{O}
(0.1\GeV)$  would be able to provide large self-interaction to alleviate the so-called small scale problems, namely ``cusp-vs-core" and ``too-big-to-fail"~\cite{Weinberg:2013aya}.

{\it Dark radiation:} $V_\mu$ and $\psi$ in the thermal bath with temperature $T_D$ will contribute as dark radiation by shifting the $N_\eff$ with
\begin{align}\label{eq:geff}
\delta  N_{\textrm{eff}}= \left(\frac{8}{7}+2\right)\left[ \frac{g_{\ast s}\left(T_\nu\right)}{g_{\ast s}\left(T^{\textrm{dec}}\right)}
\frac{g_{\ast s}^{D} \left(T^{\textrm{dec}}\right)}{g_{\ast s}^{D}\left(T_{D}\right)}
\right]^{\frac{4}{3}},
\end{align}
where $T_\nu$ is neutrino's temperature, $T^{\textrm{dec}}$ for the temperature at which dark sector is kinetically decoupled from standard model thermal bath, $g_{\ast s}$ counts the effective number of degrees of freedom (dof) for entropy density in standard model~\cite{PDG:2014}, or particles that are in kinetic equilibrium with neutrinos, $g_{\ast s}^{D}$ denotes the effective number of dof that are in kinetic equilibrium with $V_\mu$. 

Note that $\psi$ can be in thermal equilibrium with $V_\mu$ and $\Phi$ at high temperature 
due to the Yukawa interactions because it can leads to an interacting rate $\Gamma_\psi\propto g^2_\psi T^5/m_\Phi^4$. However, the gauge interaction gives rise to $\Gamma_\psi\propto g^2_XT$ and $g_X$ could be too small to keep $\psi$ in equilibrium with $V_\mu$ and $\Phi$ at high temperature, which results in a smaller $\delta N_{\textrm{eff}}$ as discussed in Ref.~\cite{Lesgourgues:2015wza}.

The above formula, Eq.~\ref{eq:geff}, is valid in general contexts. In the literature, the factor $g_{\ast s}^{D} \left(T^{\textrm{dec}}\right)/g_{\ast s}^{D}\left(T_{D}\right)$ in the bracket is usually ignored, 
which simply neglects the possible changes of dof in the dark sector. However, as shown above, 
this  ignorance is valid only if $g_{\ast s}^{D} \left(T^{\textrm{dec}}\right)\simeq g_{\ast s}^{D}
\left(T_{D}\right)$ which is not always the case. 
For instance, when $T^{\textrm{dec}}\gg m_t\simeq 173 \GeV$ for $|\lambda_{\Phi H}| \sim 
10^{-6}$, we can estimate $\delta  N_{\textrm{eff}}$ at the BBN epoch as 
\begin{equation}
\delta  N_{\textrm{eff}} =\frac{22}{7}\left[ \frac{43/4}{427/4}
\frac{11}{9/2} \right]^{\frac{4}{3}}\simeq 0.53,
\end{equation}
which shows that $g_{\ast s}^{D} \left(T^{\textrm{dec}}\right)= \dfrac{22}{9}g_{\ast s}^{D}\left(T_{D}\right)$ in our case. The lower bound can be obtained $\delta  N_{\textrm{eff}} \simeq 0.21$ when $g_{\ast s}^{D} \left(T^{\textrm{dec}}\right)=g_{\ast s}^{D}\left(T_{D}\right)$. 

We can also get the temperature ratio for $V_\mu$ to that of neutrino $\nu$ and photon $\gamma$,
\begin{equation}\label{eq:Tratio}
	T_D\simeq 0.64T_\nu=0.46T_\gamma,
\end{equation}
where we have used $T_\nu=(4/11)^\frac{1}{3}T_\gamma$. 

Based on the above discussion, the total $\delta  N_{\textrm{eff}} $ in our model is predicted to be around $0.5$, which lies in the preferred range for $\delta N_\eff \simeq 0.4-1$ to resolving the conflict between \planck ~and HST data~\cite{Riess:2016jrr}. One prediction of our model is that $\delta  N_{\textrm{eff}}>0.21$ which can be definitely either confirmed or excluded by next-generation CMB experiments.

{\it $\chi$-$\psi$ (DM-DR) scattering}: One of the key quantities for the structure formation is the 
elastic scattering cross section for $\chi + \psi \rightarrow \chi + \psi$, which would modify the 
cosmological evolutions for $\chi$ and $\psi$'s perturbations.  
More explicitly, similarly to the baryon-photon system~\cite{Ma:1995ey}, the Euler equations 
for $\chi$ and $\psi$  would be approximately modified to 
\begin{align}
\dot{\theta}_\chi &= k^2\Psi -\mathcal{H} \theta_\chi + S^{-1}\dot{\mu}\left(\theta_\psi-\theta_{\chi}\right),\label{eq:veldiv1}\\
\dot{\theta}_\psi &= k^2 \Psi +k^2\left(\frac{1}{4}\delta_\psi -\sigma_\psi\right)-\dot{\mu}\left(\theta_\psi-\theta_{\chi}\right),\label{eq:veldiv2}
\end{align}
where dot means derivative over conformal time $d\tau\equiv dt/a$ ( $a$ is the scale factor),  
$\theta_\psi$ and $\theta_\chi$ are velocity divergences of radiation $\psi$ and DM $\chi$'s, 
$k$ is the comoving wave number, $\Psi$ is the gravitational potential, $\delta_\psi$ and 
$\sigma_\psi$ are the density perturbation and the anisotropic stress potential of $\psi$, and 
$\mathcal{H} \equiv \dot{a}/a$ is the conformal Hubble parameter. Finally, the scattering rate 
and the density ratio are defined by $\dot{\mu}=a n_{\chi}\langle\sigma_{\chi \psi}c\rangle$
and $S=3\rho_\chi/4\rho_\psi$, respectively.

The averaged cross section $\langle\sigma_{\chi \psi}\rangle$ can be estimated from the squared matrix element for $\chi\psi \rightarrow \chi\psi$:
\begin{equation}\label{eq:me}
\overline{\left|\mathcal{M}\right|^{2}}\equiv \frac{1}{4}\sum_{\textrm{pol}}\left|\mathcal{M}\right|^{2} = \frac{2g_{X}^4}{t^{2}} 
\left[t^{2} + 2st + 8 m_{\chi}^{2}E_{\psi}^{2}\right],
\end{equation}
where the Mandelstam variables are $t=2E_{\psi}^{2}\left(\cos\theta-1\right)$ and $s=m_{\chi}^{2}+2m_{\chi}E_{\psi}$, where $\theta$ is the scattering angle, and $E_\psi$ is the energy of 
incoming $\psi$ in the rest frame of $\chi$. 
Integrated with a temperature-dependent Fermi-Dirac distribution for $E_\psi$, we find that 
$\langle\sigma_{\chi \psi}\rangle$ goes roughly as $g_X^4/(4\pi T^2_D)$. 

One key feature is that $\langle\sigma_{\chi \psi}\rangle$ is actually increasing as the universe is expanding, which provides a mechanism to affect the matter power spectrum ($k\gtrsim 0.1 h$/Mpc). And due to the temperature dependences of $\mathcal{H}/a\sim T^2$ at radiation-dominant era, $\mathcal{H}/a\sim T^{3/2}$ in matter-dominant era and $S^{-1}\dot{\mu}/a\sim T^2$, the last term in Eq.~7 could be equally important as $\mathcal{H}\theta_\chi$  and affect all those scales that enter horizon during radiation-dominant epoch. This is achieved because of the massless mediator, the dark photon. In the previous studies,  only the cases for 
$\langle\sigma_{\chi \psi}\rangle \propto T^2$ or $\langle\sigma_{\chi \psi}\rangle \propto$ constant 
have been widely investigated~\cite{Boehm:2014ksa,Bringmann:2013vra,Ko:2014bka}, which would 
only affect matter power spectrum at small scales. More interestingly, the mediator can also be the scattered radiation for non-abelian gauge boson~\cite{Lesgourgues:2015wza} or scalar~\cite{Tang:2016mot}, which can have very different temperature dependence and change even large scale structures (see also Refs.~\cite{Binder:2016pnr,Bringmann:2016ilk} for general discussions).  

The elastic scattering of $\chi + V_\mu \rightarrow \chi+ V_\mu$, which is similar to Compton scattering $e^- +\gamma \rightarrow e^- +\gamma $, is highly suppressed as its cross section is proportional to $g^4_X/m^2_\chi$, in comparison with $g^4_X/E^2_\psi$ for $\chi$-$\psi$ scattering. Unless $m_\chi$ is relatively light, say around $\MeV$ scale, $\chi$-$V_\mu$ scattering can be neglected through our discussion of late universe. 

\section{Numerical Results}\label{sec:numerical}

We have modified the public Boltzmann code {\tt CLASS}~\cite{class} and implemented the above 
equations, Eqs.~(\ref{eq:veldiv1}) and (\ref{eq:veldiv2}). 
We treat dark radiation $\psi$ as perfect fluid with anisotropic stress $\sigma_\psi=0$ since $\psi$'s self-interaction rate $ \sim g^4 T$ is much larger than $\mathcal{H}$ at the low temperature we are interested in.
The modification of $\theta_\chi$'s 
evolution has an impact on $\chi$'s density perturbation through 
\begin{equation}
  \dot{\delta}_\chi =  -\theta_\chi +3\dot{\Phi},
\end{equation}
where $\Phi$ is the scalar perturbation in the metric within conformal Newtonian gauge. We shall show that the interaction between DM and DR cause suppression in the matter power spectrum through diffusion damping~\cite{Boehm:2004th, Green:2005fa, Loeb:2005pm}.

\begin{figure}[t]
	\includegraphics[width=0.45\textwidth,height=0.42\textwidth]{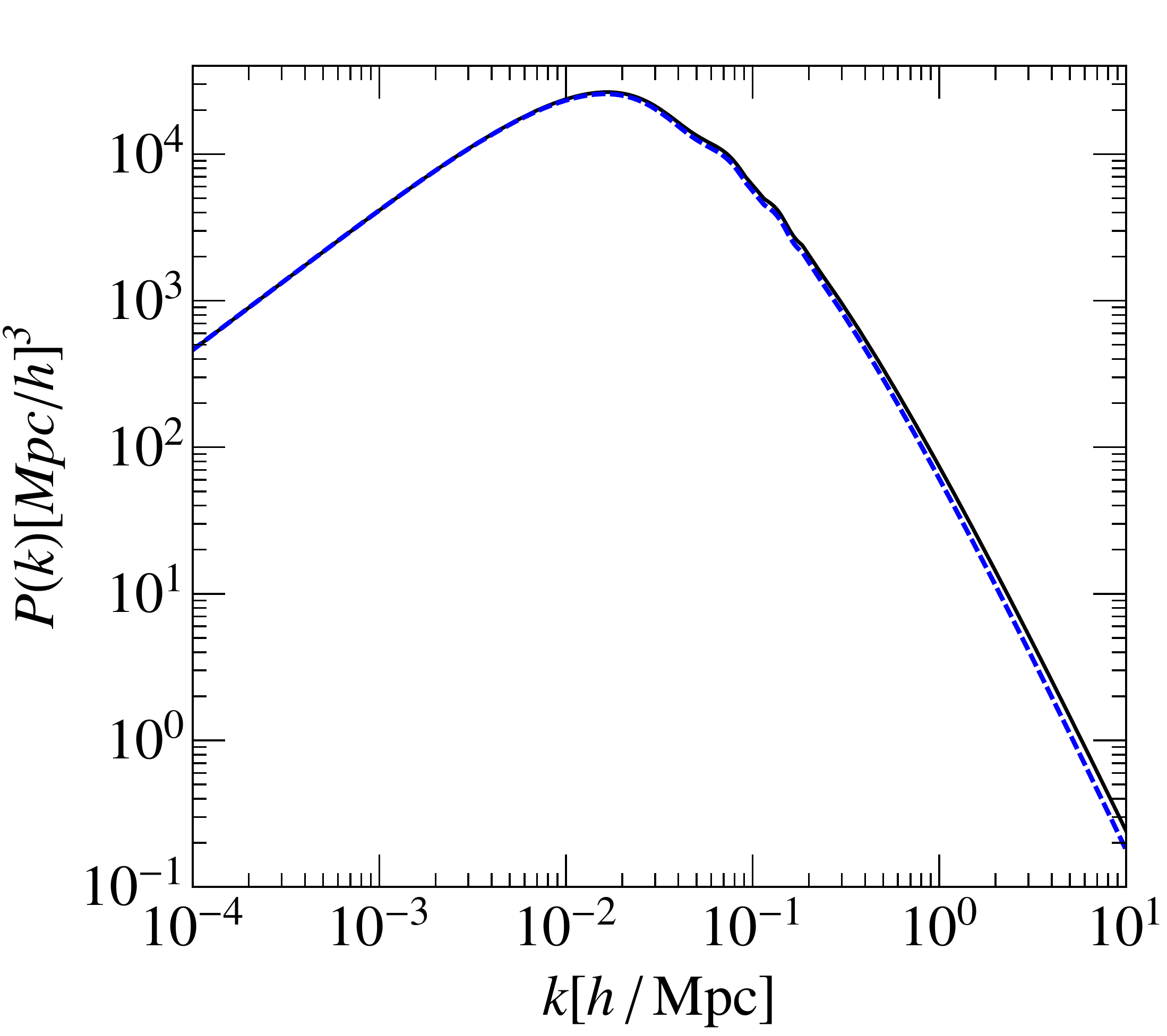}
	\includegraphics[width=0.45\textwidth,height=0.42\textwidth]{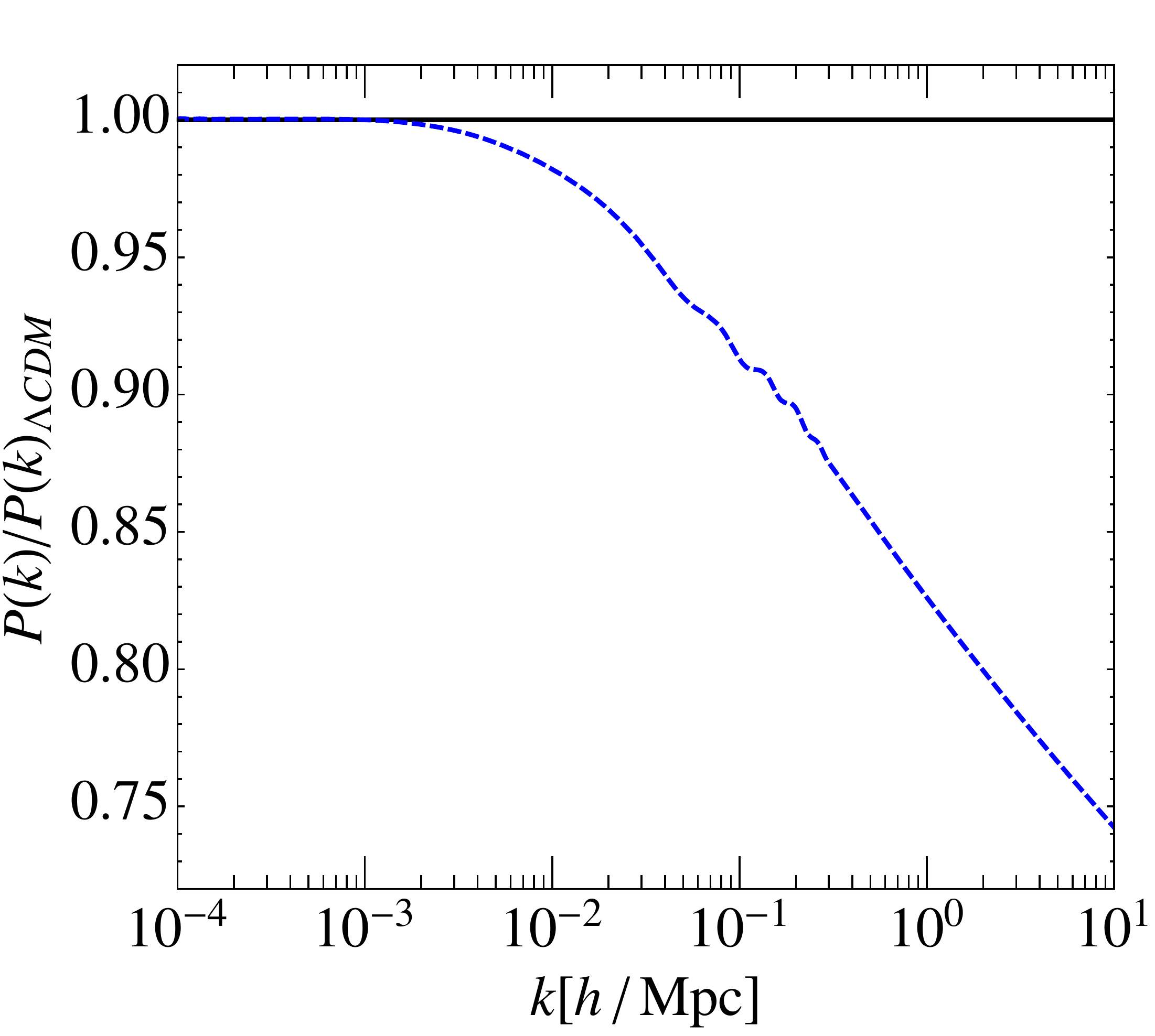}
	\caption{Illustration on matter power spectrum $P(k)$ with $m_\chi\simeq 100\GeV$ and $g^2_X\simeq 10^{-8}$. The black solid lines are for $\Lambda$CDM and the blue dashed lines for interacting DM-DR case. We can easily see that $P(k)$ is suppressed for modes that enter horizon at radiation-dominant era. }
	\label{fig:sigma8}
\end{figure}

We illustrate the physical effect in Fig.~\ref{fig:sigma8}. The upper panel shows the matter power spectrum $P(k)$, solid (dashed) line for $\Lambda$CDM (interacting DM) case, and the lower panel shows the ratio. We have chosen $m_\chi\simeq 100\GeV$ and $g_X^2\simeq 10^{-8}$. It can be clearly seen that the matter power spectrum is suppressed, therefore gives a smaller $\sigma_8$. For the parameters we used, the suppression is about $10\%$ at $k\simeq h/8\textrm{Mpc}$, enough for relaxing the tension between \planck ~and weak lensing data. Unlike the scenarios~\cite{Cyr-Racine:2013fsa,Bringmann:2013vra,Ko:2014bka,Chu:2014lja} where DM-DR scattering $\langle\sigma_{\chi \psi}\rangle$ has positive-power dependence on the temperature, this model has negative-power dependence and predicts smooth suppression.

We take the central values of six parameters of $\Lambda$CDM from $\planck$~\cite{Planck:2015xua},
\begin{align}
&\Omega_b h^2=0.02227, \Omega _c h^2 =0.1184, 100\theta_{\textrm{MC}}=1.04106, \nonumber \\
&\tau = 0.067, \ln\left(10^{10}A_s\right)=3.064, n_s=0.9681,
\end{align} 
which gives $\sigma_8=0.817$ in vanilla $\Lambda$CDM cosmology. With the same input as above, now we take $\delta N_\eff \simeq 0.53$,  $m_\chi\simeq 100\GeV$ and $g_X^2\simeq 10^{-8}$ in the interacting DM case, we have $\sigma_8\simeq 0.744$ which is much closer to the value $\sigma_8\simeq 0.730$ given by weak lensing survey CFHTLenS~\cite{Heymans:2012gg}.

Dedicated analysis with Markov Chain Monte Carlo for statistical inference of the precise 
parameters is beyond our scope in this paper. However, we can understand the  physics of  
collisional damping and roughly estimate the size of $g_X$ by comparing $\mathcal{H}$ and 
$ S^{-1}\dot{\mu}$ in Eq.~(\ref{eq:veldiv1}),
\begin{equation}\label{eq:crit}
\frac{S^{-1}\dot{\mu}}{\mathcal{H}}=\frac{S^{-1}n_\chi\langle \sigma_{\chi \psi }c\rangle}{\mathcal{H}/a}\sim \frac{T_D n_\psi\langle\sigma_{\chi \psi }c\rangle}{m_\chi H}\gtrsim 1,
\end{equation}
where the Hubble parameter $H$ is given by $T^2/M_{pl}$ ($M_{pl}\simeq 10^{18}\GeV$) in radiation-dominant era. Requiring the above inequality hold until matter-dominant time, we can obtain
\begin{equation}
g^2_X \sim \frac{T_\gamma}{T_D}\left(\frac{m_\chi}{M_{pl}}\right)^{1/2}.
\end{equation}
Since $T_\gamma/T_D\sim 2$ as shown in Eq.~(\ref{eq:Tratio}), we would have 
$g_X^2\sim 10^{-8}$  for $m_\chi\simeq 100\GeV$. It is also evident that increasing DM mass 
$m_\chi$ or deceasing DR temperature $T_D$ would require large $g_X$.

From the above discussions, it is also clear that the new gauge boson $V_\mu$ does not have to be strictly massless. As long as its mass is much smaller than temperature $T_D$ around radiation-matter equality time, say $m_V \ll 0.1\eV$, our above discussions still hold. 
This can be easily achieved if the scalar $\Phi$ develops nonzero but tiny VEV,
or if the dark photon gets massive by nonzero VEV of another $U(1)_X$-charged scalar with 
$U(1)_X$ charge different from $\Phi$'s. 
There might be a slight change since $V_\mu$ then would decay into $\psi$ pairs and modify 
the number of $\psi$ in Eq.~(\ref{eq:crit}). Also the roles played by scalar and vector can be interchanged, namely scalar mediates DM-DR interaction and vector is responsible for the relic density.  

Likewise the fermionic DR $\psi$ needs not be strictly massless, and could get tiny mass $\lesssim 0.1\eV$ to be still relativistic around the radiation-matter equality time. Then it would behave as a light sterile neutrino with dark 
interaction, which is still allowed by astrophysics or cosmology as long as the mixing with 
active neutrinos is small enough. DR can also be bosonic, see Refs.~\cite{Lesgourgues:2015wza, Tang:2016mot, Jeong:2013eza} for scalar and vector boson as examples.

The above mechanism can work for other DM-DR models as well.
For example, dark matter can be a complex scalar $X$ rather than a Dirac fermion. 
Here we present a local $Z_3$ scalar DM model originating from dark $U(1)_X$ gauge symmetry 
\cite{Ko:2014nha,Ko:2014loa}, 
in which the dark Higgs $\phi_X$ has a dark charge $3$ while DM $X$ has a dark charge $1$.
Then the renormalizable Lagrangian involving these new fields is given by  
\begin{align} \label{eq:z3}
{\cal L} = &D_{\mu}X^{\dagger}D^{\mu}X+\bar{\psi}i\slashed{D}\psi-\frac{1}{4}V_{\mu\nu}V^{\mu\nu}+D_{\mu}\phi_{X}^{\dagger}D^{\mu}\phi_{X}-V,
\end{align}
where the scalar potential $V$ is given by
\begin{align}
V = & -\mu_{\phi}^{2}\phi_{X}^{\dagger}\phi_{X}+\lambda_{\phi}\left(\phi_{X}^{\dagger}\phi_{X}\right)^{2}+\mu_{X}^{2}X^{\dagger}X+\lambda_{X}\left(X^{\dagger}X\right)^{2}  \nonumber \\
+ & \lambda_{\phi H}\phi_{X}^{\dagger}\phi_{X}H^{\dagger}H+\lambda_{\phi X}X^{\dagger}X\phi_{X}^{\dagger}\phi_{X}+\lambda_{HX}X^{\dagger}XH^{\dagger}H  \nonumber \\
+ & \left( \lambda_{3}X^{3}\phi_{X}^{\dagger}+H.c. \right), 
\end{align}
where $H$ is the SM Higgs doublet. After $\phi_X$ gets a small VEV, we have a cubic term 
$X^3$ with $Z_3$ symmetry which protects $X$'s stability even in the presence of 
nonrenormalizable higher dimensional operators.  $X$-$\psi$'s scattering and other effects are 
similar to what we discussed above, except that now new Higgs-portal term can provide direct detection signals.

\section{Discussion}

Besides the thermal history and $\delta N_{\eff}$, let us know now discuss some other differences from the $U(1)$ scenario sketched briefly in Ref.~\cite{Lesgourgues:2015wza} which is actually mostly devoted to electroweak-charged DM with hidden non-Abelian gauge interaction. Based on what we understand from Ref.~\cite{Lesgourgues:2015wza}, we list some differences below:

1. The Dirac DM candidate in Ref.~\cite{Lesgourgues:2015wza} is a chiral fermion, so it is necessary to introduce other chiral fermions to cancel the gauge anomalies. It then can be interpreted that the model presented in Ref.~\cite{Lesgourgues:2015wza} is an effective theory. In our proposal, however, the DM candidate is vector-like, so the theory is automatically anomaly-free and therefore can be an ultraviolet complete model.  

2. Due to the electroweak interaction of DM particle in Ref.~\cite{Lesgourgues:2015wza}, the indirect searches also actually put stringent constraints on the mass of DM $\gtrsim \mathcal{O}(\TeV)$ due to the gamma-rays from the annihilation of DM into electroweak bosons. While in our model, the dominant channel is $\chi+\bar{\chi}\rightarrow \Phi +\Phi^\dagger$, followed by $\Phi$'s decay into dark radiation $\psi$ and right-handed neutrino $N$. $N$ can mix with and oscillate into left-handed neutrino $\nu_a$. Since the current constraint from IceCube neutrino searches is much weaker than gamma-ray's limit, the range of DM's mass in our model can be significantly larger. 

3. One more difference is about exotic decay of SM Higgs $h$. If the scalar $\Phi$'s mass in our model is less than $M_h/2\simeq 62.5\GeV$, the SM Higgs boson $h$ can decay into $\Phi + \Phi^\dagger$ and give invisible decay channel of $h$. The current limit can actually constrain $\lambda_{\Phi H}\lesssim 10^{-3}$. While in Ref.~\cite{Lesgourgues:2015wza}, no invisible decay channel is expected.

\section{Summary}\label{sec:summary}
In this paper, we have investigated a model for a dark sector where dark matter (DM) interacts 
with fermionic dark radiation (DR) through a light gauge boson (dark photon) 
in order to resolve some tensions in cosmological data. 
The new light gauge boson (dark photon) plays a key role both in the DM-DR 
elastic  scattering and in the late kinetic decoupling. This simple model can provide the right 
amount of  DR ($\delta N_\eff \sim 0.5$), thereby resolving the tension in Hubble constant $H_0$ 
between \planck ~and HST data.  Also the elastic scattering between DM and DR causes the 
collisional damping and has impact on the structure growth rate, which leads to a smaller 
$\sigma_8$ and relaxes the conflicts between \planck ~ and weak lensing measurement. 
Finally the light fermionic DR $\psi$ can be interpreted as a sterile neutrino in some models.
And all these niceties rely on the underlying local dark gauge symmetry. 

\begin{acknowledgments}
This work is supported in part by National Research Foundation of Korea 
(NRF) Research Grant NRF-2015R1A2A1A05001869 (PK,YT), and by the NRF grant 
funded by the Korea government (MSIP) (No. 2009-0083526) through Korea 
Neutrino Research Center at Seoul National University (PK).
\end{acknowledgments}


\providecommand{\href}[2]{#2}\begingroup\raggedright\endgroup

\end{document}